\documentclass[prl,twocolumn,showpacs,preprintnumbers,amsmath,amssymb]{revtex4}
\usepackage{graphicx}
\usepackage{dcolumn}
\usepackage{bm}
\usepackage{psfig}
\newcommand \be{\begin{eqnarray}}
\newcommand \ee{\end{eqnarray}}

\begin{document}
\title{The chemical potential for the inhomogeneous electron liquid in terms of its kinetic and potential parts with special consideration of the surface potential step and BCS-BEC crossover}
\author{K. Morawetz$^{\ast}$$^{1,2}$\thanks{$^\ast$Corresponding author. Email: morawetz@physik.tu-chemnitz.de
\vspace{6pt}}N. H. March${^{3,4,5}}$ and R. H. Squire$^6$}
\affiliation{$^1$ Institute of Physics, Chemnitz University of Technology, 
09107 Chemnitz, Germany}
\affiliation{
$^2$ Max-Planck-Institute for the Physics of Complex
Systems, N{\"o}thnitzer Str. 38, 01187 Dresden, Germany}
\affiliation{
$^3$Department of Physics, University of Antwerp, Belgium}
\affiliation{
$^4$Oxford University, Oxford, England}
\affiliation{
$^5$Abdus Salam International Centre for Theoretical Physics, Trieste, Italy}
\affiliation{
$^6$Department of Chemistry, West Virginia University, Montgomery, WV25136, USA
}
\begin{abstract}
The chemical potential $\mu$ of a many-body system is valuable since it carries fingerprints of phase changes. Here, we summarize results for $\mu$ for a three-dimensional electron liquid in terms of average kinetic and potential energies per particle. The difference between $\mu$ and the energy per particle is found to be exactly the electrostatic potential step at the surface. We also present calculations for an integrable one-dimensional many-body system with delta function interactions, exhibiting a BCS-BEC crossover. It is shown that in the BCS regime the chemical potential can be expressed solely in terms of the ground-state energy per particle. A brief discussion is also included of the strong coupling BEC limit. 
\end{abstract}
\date{\today}
\maketitle

The dependence of the chemical potential $\mu$ at $T=0$ on a physical observable such as particle number density is considered for two quantum liquids. As a starting point, the homogeneous electron liquid in three dimensions is discussed by combining the quantum-mechanical virial theorem with results of Hugenholtz and Van Hove and the Budd-Vannimenus theorem. As the electron density is lowered, $\mu$ is related to surface properties. 

Secondly, consideration is given to a recent one-dimensional model possessing a crossover from a Bardeen-Cooper-Schrieffer (BCS) state to a Bose-Einstein condensate (BEC). This crossover occurs at $\mu=0$, changing from positive values of $\mu$ in the BCS phase through zero into the BEC phase at negative $\mu$. Using the thermodynamic relations like $\mu=d E/dN$ where $E$ is the ground-state energy and N is the number of particles, $E$ and $\mu$ are readily related in the one-dimensional model.

One of us \cite{M72} in early work combined a many-body result of Hugenholtz and Van Hove \cite{HH58} with the virial theorem \cite{Ma58,A67} for a homogeneous correlated electron liquid. This led to an expression for the chemical potential $\mu$ as a linear combination of the average kinetic energy $\langle T \rangle $, and the corresponding average potential energy $\langle V\rangle  $
\be
N \mu=\frac 5 3 \langle T\rangle  +\frac 4 3 \langle V\rangle 
\label{m}
\ee 
where $N$ denotes the number of electrons. Of course $\langle T\rangle  $ and $\langle V\rangle  $ depend on the value of the uniform electron density $n$ or equivalently on the mean inter-electronic spacing $r_s$
\be
{4 \pi\over 3} r_s^3={V \over N}={1\over n}.
\label{rs}
\ee
One knows from free Fermi gas theory that as $r_s\to0$, i.e. in the extreme high-density limit, the kinetic energy per electron is $3/5$ of the Fermi energy, which in this limit is the chemical potential. Thus one can obtain insight as $r_s\to0$ into the exact result (\ref{m}) for $\mu$ in the correlated uniform electron liquid of density $n$. The quantum-mechanical averages of the kinetic and potential energies in (\ref{m}) respectively depend on $n$ or by (\ref{rs}) on $r_s$. As $r_s\to0$, the kinetic energy per particle $\langle T\rangle  /N$ is proportional to $r_s^{-2}$ and the potential energy per particle $\langle V\rangle  /N$ to $r_s^{-1}$ and hence the kinetic energy dominates the potential term, as required in the free Fermi gas limit. Consequently, as $r_s$ is increased, the chemical potential $\mu$ will pass from a large positive value at high density, through zero, to a negative value at sufficiently low density.

It is now interesting to note that the exact relation (\ref{m}) can be rewritten into another form revealing the link to surface properties. Therefore we repeat the steps leading to (\ref{m}) but with the explicit inclusion of the surface assuming for simplicity that we have a flat surface and the electron liquid in the half space $x<0$. Then the Budd-Vannimenus (BV) theorem \cite{BV73} links the surface potential step of the extended electron system directly to the bulk energy $E_b/N$ per particle
\be
\Delta \Phi = \phi(0)-\phi(-\infty)=-{r_s\over 3} {\partial \left ({E_b/ N}\right ) \over \partial r_s}.
\label{BV}
\ee
The electrons leak out of the surface which leads to a depletion of electron density inside the electron system such that the surface dipole develops that results in the electrostatic potential step (\ref{BV}). This BV theorem for special geometries can be found in \cite{ZL83} and has played an important role to clarify discrepancies in observations of thermodynamic corrections in the Bernoulli potential at superconducting surfaces \cite{LMKMBSb04, LKMB02}. 

Compared to the original form \cite{Ma58}, the virial theorem now contains an additional surface term \cite{KiWo96}
\be
2 {\langle T\rangle \over N}+{\langle V \rangle \over N}=-r_s {\partial \left ({E_b/N}\right )\over \partial r_s}- {A\over N}\left (r_s {\partial \over \partial r_s}+2 \right ) \sigma
\label{virial}
\ee
where $A$ is the surface area and $\sigma$ the surface energy. Distinguishing the surface and bulk parts in the kinetic, $T=T_b+T_\sigma A/N$, and potential energies, $V=V_b+V_\sigma A/N$ one obtains the virial theorems separately for bulk and surface properties, respectively \cite{KiWo96}
\be
2 \langle T_b\rangle  +\langle V_b\rangle  &=&-N r_s{\partial \left ({E_b/ N} \right )\over \partial r_s}
\nonumber\\
2 \langle T_\sigma\rangle  +\langle V_\sigma\rangle  &=&-N \left (r_s {\partial \over \partial r_s}+2\right )
{\sigma} .
\label{virial1}
\ee
With $E_b=\langle T_b\rangle  +\langle V_b\rangle  $ and $\sigma=\langle T_\sigma\rangle  +\langle V_\sigma\rangle  $ one can find the kinetic and potential parts explicitly
\be
\begin{array}{lclclcl}
{\langle T_b \rangle\over N}&=&-r_s{\partial \over \partial r_s} \left (r_s {E_b\over N} \right )
&\qquad&
{\langle V_b \rangle\over N}&=&{1\over r_s}{\partial \over \partial r_s}  \left (r_s^2 {E_b\over  N} \right )
\cr&&&&&&\cr
{\langle T_\sigma \rangle\over N}&=&-{1\over r_s^2}{\partial \over \partial r_s}\left (r_s^3 \sigma\right )
&\qquad&
{\langle V_\sigma \rangle\over N}&=&{1\over r_s^3}{\partial\over \partial r_s} \left (r_s^4 \sigma\right ).
\end{array}
\label{ve}
\ee
Introducing the pressure $p$ by 
utilizing the Hugenholtz-Van Hove theorem \cite{HH58}
\be
N\mu=E_b\!+\!pV=E_b\!-\!V\!\left (\!{\partial E_b\over \partial V} \!\right )_{S,N}\!\!\!=E_b\!-\!{r_s\over 3} \!\left (\!{\partial {E_b}\over \partial r_s}\!\right )_{S,N}
\label{m1}
\ee
and eliminating in (\ref{virial1}) the derivative of $E_b$ with the help of (\ref{m1}), we obtain (\ref{m}) for the bulk parts.

With the help of the virial theorem (\ref{virial}), the electrostatic potential step at the surface (\ref{BV}) can now be expressed in terms of the kinetic and potential bulk energy and the surface tension as
\be
N \Delta \Phi =\frac 2 3 \langle T\rangle  +\frac 1 3 \langle V\rangle+{A\over 3} \left (r_s {\partial \over \partial r_s}+2 \right ) \sigma. 
\label{BV1}
\ee
Analogously to \cite{M72} we now employ again the Hugenholtz-Van Hove theorem (\ref{m1}) and
using (\ref{BV1}) and (\ref{virial}) we obtain 
\be
N\mu=E_b+ N \Delta \Phi.
\label{mnew}
\ee
This formula is a central result of this manuscript and is not found in the literature to the authors' knowledge. It describes the chemical potential in terms of the bulk energy and the potential step at the surface. The potential step at the surface is just the difference between the chemical potential and the total energy per particle. 

For infinite (homogeneous) electron systems we can shift the surface to infinity. Thus we can now interpret the part beyond the bulk energy in the chemical potential (\ref{m}), 
\be
N \Delta\Phi&=&N \mu-{E_b}
=\frac 2 3 {\langle T_b\rangle } +\frac 1 3 {\langle V_b\rangle }\equiv {p} V,
\label{new}
\ee 
as counting a surface potential step infinitely far away. 
Eq. (\ref{new}) is exact. The identification of the surface potential step with the pressure (\ref{new}) comes from the Hugenholtz-Van Hove theorem again. It is physically clear that the surface potential step also accounts for the pressure.

Our interpretation also agrees with the expectation that the chemical potential should describe the energy necessary to add or remove a particle in the 
system. The terms beyond $E_b$ expressed in (\ref{m}) can be consequently viewed as coming from an imagined surface. Therefore adding or removing a particle requires to overcome the energy of the bulk as well as the surface potential step.

This refined picture allows us now to interpret the BCS-BEC transition as a competing effect between the bulk and surface potential step. For the weak coupling (BCS) limit the chemical potential is positive and the bulk energy $E$ exceeds the negative surface potential step $\Delta \Phi$. In the strong coupling limit (BEC) the surface part can be considered as dominant. The system is unstable to the creation of bound states which can Bose condense. 

With this brief summary and reinterpretation of the different parts of the chemical potential in the three-dimensional electron liquid, we turn now to an integrable one-dimensional system. This will allow us to discuss the above parts of the chemical potential explicitly. We will pay special attention to the different kinetic and potential parts as well as the surface potential step.

In recent work, Wadati and Iida \cite{WI07} have considered the crossover from a BCS-like state to a BEC state in a one-dimensional integrable model. The Hamiltonian assumed for $N$ particles in a periodic box of length L takes the form
\be
H=-\sum\limits_{j=1}^N {h^2\over 2 m} {\partial^2 \over \partial x_j^2}+c \sum\limits_{i\ne j} \delta(x_i-x_j),
\ee
with $m$ being the mass and $c$ the coupling constant. Following \cite{WI07} we choose units $\hbar=1$ and $2 m=1$ for simplicity. It is useful to define the number 'density' $D=N/L$ of particles and the dimensionless coupling parameter
\be
\gamma ={c\over D}.
\ee

\begin{figure}
\psfig{file=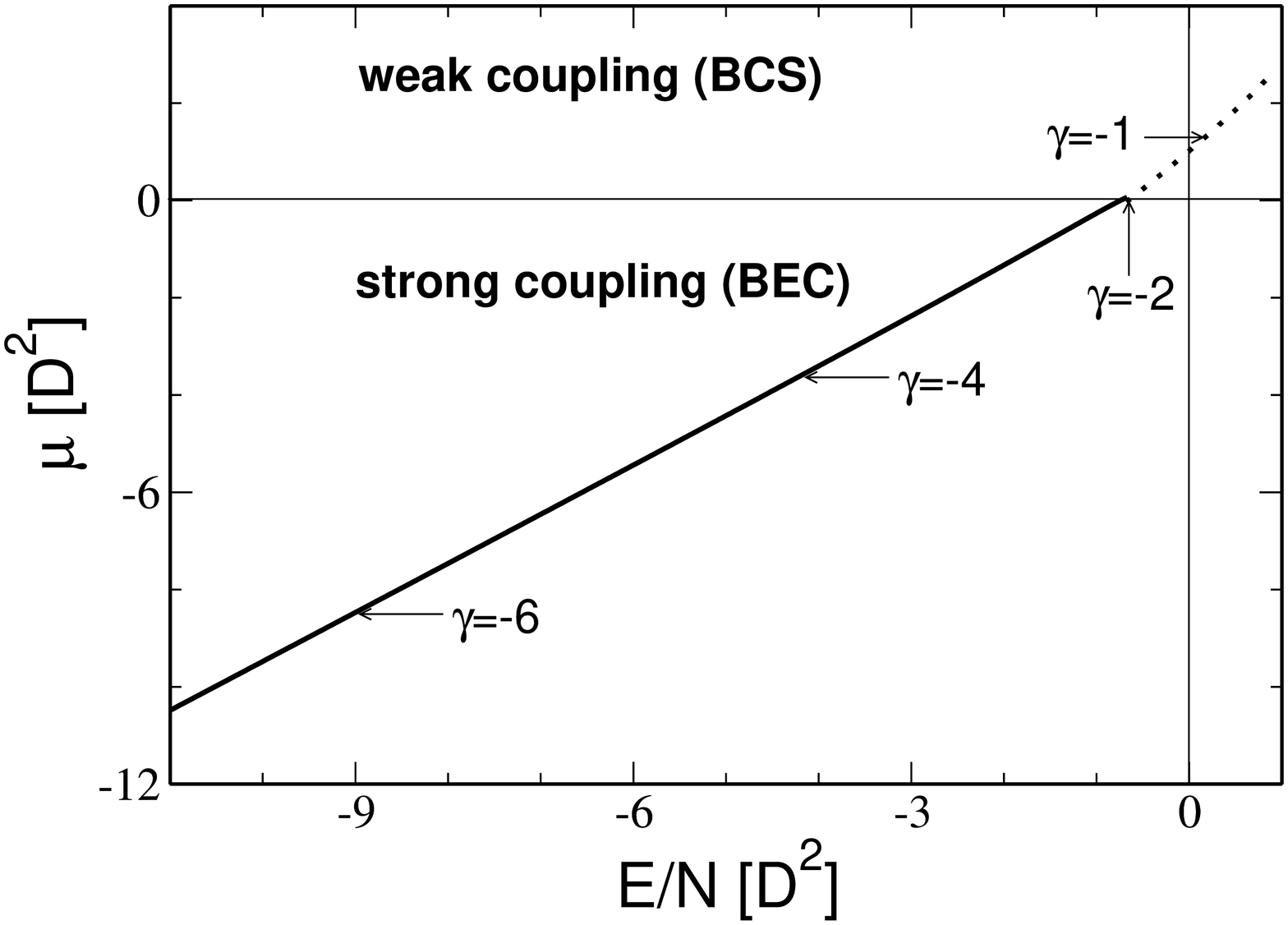,width=9cm}
\caption{The chemical potential as a function of the total energy per particle for the weak (dotted line) and strong coupling (solid line) limit.}
\label{weak}
\end{figure}

Wadati and Iida \cite{WI07} solve the so-called Gaudin integral equation, see also \cite{SM07}, in the weak coupling BCS regime corresponding to $|\gamma|\ll1$, and hence obtain the ground state energy as
\be
{E_{\rm BCS}\over N D^2}&=&{\pi^2\over 12} \left (1+{6 \gamma\over \pi^2}-{3\gamma^2\over 2 \pi^2} \right ) +{\cal O}(\gamma^3)
\label{33}
\ee 
which according to (\ref{ve}) is composed of kinetic and potential parts as
\be
{\langle T_{b}\rangle_{\rm BCS}\over N D^2}&=&{\pi^2\over 12} \left (5+{12 \gamma\over \pi^2}+{3\gamma^2\over 2 \pi^2} \right ),
\nonumber\\
{\langle V_{b}\rangle_{\rm BCS}\over N D^2}&=&{\pi^2\over 12} \left (-4-{6 \gamma\over \pi^2}-{6\gamma^2\over 2 \pi^2} \right ).
\label{vew}
\ee
We now utilize the thermodynamic relation $\mu=\partial E/ \partial N$ to get the chemical potential
\be
\mu_{\rm BCS}={D^2 \pi^2\over 4} \left (1+{4 \gamma\over \pi^2}-{\gamma^2\over 2 \pi^2} \right )+{\cal O}(\gamma^3)
\label{35}
\ee
which agrees with the result of \cite{WI07} when a misprint is corrected in their Eq. (3.25). 
From (\ref{35}) with the help of (\ref{33}) we obtain the relation
\be
\mu_{\rm BCS}={D^2 \pi^2\over 12} \left (2+{6 \gamma\over \pi^2} \right )+{E_{\rm BCS} \over N}.
\label{22}
\ee
In the limit $\gamma\to0$, it follows from (\ref{33}) that ${E_{\rm BCS} /N}\to \pi^2  D^2/12$, and from (\ref{22}) that $\mu_{\rm BCS}\to\pi^2  D^2/4$. 
Removing $\gamma$ in (\ref{33}) and (\ref{22}) yields a quadratic equation for the chemical potential in terms of the energy, namely
\be
\left ({\pi^2\over 6} +{E_{\rm BCS} \over N D^2}-+{\mu_{\rm BCS} \over D^2}\right)^2={\pi^2\over 6}(\gamma-1)+4 {E_{\rm BCS} \over N D^2}
\label{qu}
\ee 
where the physical solution is plotted in figure~\ref{weak}. We see that the weak coupling limit $\gamma\ll 2$ describes the range where $\mu>0$ and the transition from negative energy to positive energy. The latter transition corresponds to the transition of bound to free matter. The interpretation is that for $E<0$ the electrons are bound representing Bosons which can form a Bose condensation for $\mu<0$ if the density is high enough in the ground state. In order to describe this strong coupling regime, we have to consider the other limiting result.

The scaled ground state energy in the strong coupling regime reads \cite{WI07}
\be
{E_{\rm strong}\over N D^2}=-\frac 1 4 \gamma^2 +{\pi^2\over 12} \left ({\gamma\over 2 \gamma\!+\!1}\right )^2 \left (1 \!+\!{4 \pi^2\over 15 (2 \gamma\!+\!1)^3} \right )
\label{18}
\ee
which again from(\ref{ve}) is composed of the kinetic and potential parts
\be
{\langle T_{b}\rangle_{\rm strong}\over N D^2}&=&
\frac{\gamma^2}{4}+{\pi^2\over 12} \left (\frac{\gamma}{1+2 \gamma}\right )^2
\nonumber\\&&\times\left [
5+\frac{6}{1+2 \gamma} \left(-1+\frac{2 (-1+28 \gamma) \pi
 ^2}{45 (1+2 \gamma)^3}\right)\right]
\nonumber\\
{\langle V_{b}\rangle_{\rm strong}\over N D^2}&=&-\frac{\gamma^2}{2}+{\pi^2\over 12 }\left (\frac{\gamma}{1+2 \gamma}\right )^2 
\nonumber\\&&\times
\left[-4\!+\!\frac{6 }{1\!+\!2 \gamma}\left(1\!+\!\frac{2 (2\!-\!26 \gamma) \pi
 ^2}{45 (1\!+\!2 \gamma)^3}\right)\right].
\label{ves}
\ee
From this we obtain the chemical potential observing that $D=N/L$ and $\gamma=c L/N$
\be
{\mu_{\rm strong}\over D^2}&=&-\frac 1 4 \gamma^2+{\pi^2\over 12} {6 \gamma+1\over 2 \gamma+1} \left ({\gamma\over 2 \gamma+1} \right )^2 
\nonumber\\&&\times\left ( 1+{4 \pi^2 (1+12 \gamma)\over 15 (1+2 \gamma)^3 (1+6\gamma)} \right )  
\ee
which corrects some misprint in the last term of Eq. (3.8) in \cite{WI07}.
The curves $\mu(\gamma)$ and $E_b(\gamma)$ fit smoothly at $\gamma=-2$. We see also that this corresponds nearly to the point where the chemical potential changes sign which identifies with the BCS-BEC transition point, exactly being $\gamma_{\rm BCS}^c \approx-1.98$ and $\gamma_{\rm strong}^c\approx -2.07$. 

\begin{figure}
\psfig{file=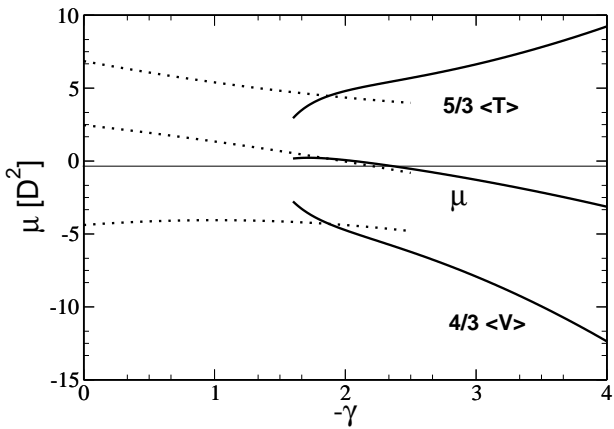,width=9cm}
\caption{The chemical potential and its kinetic and potential parts as a function of the coupling parameter for the weak (dotted line) and strong-coupling (solid line) limit.}
\label{mvt}
\end{figure}

In figure~\ref{mvt} we compare the kinetic and potential contribution to the chemical potential according to (\ref{m}) calculated in (\ref{ves}) and (\ref{vew}). We see that though the resulting chemical potential is smooth at the transition point $\gamma\approx 2$ the kinetic and potential energies possess different slopes for the BEC and BCS sides. This points towards the different correlation mechanisms in both regimes.

\begin{figure}
\psfig{file=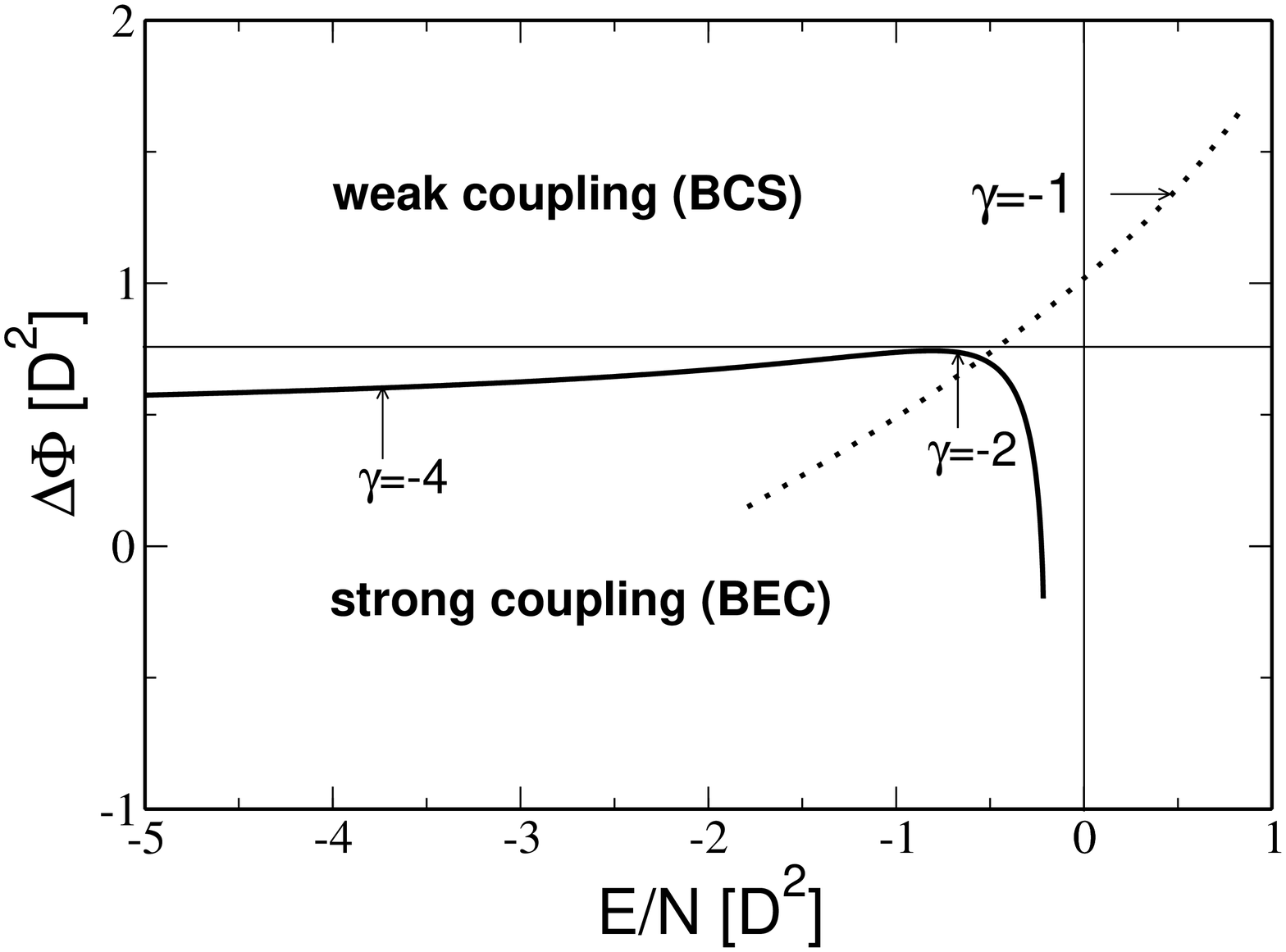,width=9cm}
\caption{The surface potential step as a function of the total energy per particle for the weak (dotted line) and strong-coupling (solid line) limit. The formula (\protect\ref{18}) of \cite{WI07} is not applicable for $\gamma>-2$.}
\label{surfp}
\end{figure}

Next we investigate the surface potential step due to the new formula (\ref{new}). We obtain the weak and strong-coupling limits
\be
\Delta \Phi_{\rm BCS}&=&D^2{\pi^2\over 6}\left (1+{3\over \pi^2} \gamma \right )+{\cal O}(\gamma^3)\nonumber\\
\Delta \Phi_{\rm strong}&=&D^2{\pi^2\over 3} \left ({\gamma\over 1\!+\!2 \gamma}\right )^3\left (1\!+\!{2\pi^2\over 3(1\!+\!2\gamma^2)^3}\right )  
\ee
which are plotted in figure~\ref{surfp}.  
The surface potential step shows a sharp change in the slope at the BCS-BEC transition point. While for the strong-coupling limit the surface potential step is only growing slowly with the energy it rises sharply in the weak coupling limit. The divergence of the strong-coupling limit for $\gamma>-2$ is present in the above formulas for the energy and chemical potential as well and is attributed to the approximation (\ref{18}) used in (\ref{new}) where $\gamma>-2$ exceeds the strong-coupling regime.

The chemical potential $\mu$ in many-body systems contains fingerprints of phase changes. For the three-dimensional correlated homogeneous electron liquid, $\mu$ is given exactly in terms of the average kinetic and potential energies for electrons in eqn. (\ref{m}). The high-density limit $r_s\to0$ is dominated by the kinetic energy. However, as the density $n$ is lowered, corresponding to increasing $r_s$, $\mu$ passes through zero, which is related to surface phenomena. Utilizing the Budd-Vannimenus theorem we have been able to identify the difference between the chemical potential and the total energy per particle as the surface potential step assuming that we have an infinite flat surface of half space. This allows us to view the sign change of the chemical potential as a competing effect of total energy and surface potential step.

Though we have restricted our discussion to a uniform electron liquid, for $\mu$ sufficiently negative the liquid is unstable against Wigner electron crystallization \cite{CM75}.

Turning finally to the one-dimensional model of Wadati and Iida \cite{WI07} possessing a BCS-BEC crossover, it has been demonstrated that the chemical potential $\mu$ can be expressed solely in terms of the ground-state energy per particle. The coupling constant of the electrons in the many body Hamiltonian no longer appears explicitly. We calculated the surface potential step according to our finding (\ref{new}) and observed that the BCS-BEC transition is characterized by a change in the slope of the surface potential step with the ground state energy per particle. This could be viewed as an alternative measure for the BCS-BEC transition.

It would, of course, be interesting if an analytically tractable model in higher dimensions could be devised in which $\mu=\mu(E/N)$ could be obtained, showing a BCS-BEC crossover.

K.M. and N.H.M. wish to acknowledge that their contribution to this Letter resulted from participation in an International Workshop at the Maths Centre of SNS Pisa in 2007. They thank the Centre both for the very stimulating atmosphere and for generous hospitality. N.H.M. and R.H.S. acknowledge that their collaboration resulted from the Sanibel Symposium (2007) in which they both participated.

\bibliography{sem3,bose,kmsr,kmsr1,kmsr2,kmsr3,kmsr4,kmsr5,kmsr6,kmsr7,delay2,spin,refer,delay3,gdr,chaos,sem1,sem2,short,genn}

\begin{thebibliography}{12}
\expandafter\ifx\csname natexlab\endcsname\relax\def\natexlab#1{#1}\fi
\expandafter\ifx\csname bibnamefont\endcsname\relax
  \def\bibnamefont#1{#1}\fi
\expandafter\ifx\csname bibfnamefont\endcsname\relax
  \def\bibfnamefont#1{#1}\fi
\expandafter\ifx\csname citenamefont\endcsname\relax
  \def\citenamefont#1{#1}\fi
\expandafter\ifx\csname url\endcsname\relax
  \def\url#1{\texttt{#1}}\fi
\expandafter\ifx\csname urlprefix\endcsname\relax\def\urlprefix{URL }\fi
\providecommand{\bibinfo}[2]{#2}
\providecommand{\eprint}[2][]{\url{#2}}

\bibitem[{\citenamefont{March}(1972)}]{M72}
\bibinfo{author}{\bibfnamefont{N.~H.} \bibnamefont{March}},
  \bibinfo{journal}{Phys. Lett. A} \textbf{\bibinfo{volume}{39}},
  \bibinfo{pages}{150} (\bibinfo{year}{1972}).

\bibitem[{\citenamefont{Hugenholtz and {Van~Hove}}(1958)}]{HH58}
\bibinfo{author}{\bibfnamefont{N.~M.} \bibnamefont{Hugenholtz}}
  \bibnamefont{and}
  \bibinfo{author}{\bibfnamefont{L.}~\bibnamefont{{Van~Hove}}},
  \bibinfo{journal}{Physica} \textbf{\bibinfo{volume}{24}},
  \bibinfo{pages}{363} (\bibinfo{year}{1958}).

\bibitem[{\citenamefont{March}(1958)}]{Ma58}
\bibinfo{author}{\bibfnamefont{N.~H.} \bibnamefont{March}},
  \bibinfo{journal}{Phys. Rev.} \textbf{\bibinfo{volume}{110}},
  \bibinfo{pages}{604} (\bibinfo{year}{1958}).

\bibitem[{\citenamefont{Argyres}(1967)}]{A67}
\bibinfo{author}{\bibfnamefont{P.~N.} \bibnamefont{Argyres}},
  \bibinfo{journal}{Phys. Rev.} \textbf{\bibinfo{volume}{154}},
  \bibinfo{pages}{410} (\bibinfo{year}{1967}).

\bibitem[{\citenamefont{Budd and Vannimenus}(1973)}]{BV73}
\bibinfo{author}{\bibfnamefont{H.~F.} \bibnamefont{Budd}} \bibnamefont{and}
  \bibinfo{author}{\bibfnamefont{J.}~\bibnamefont{Vannimenus}},
  \bibinfo{journal}{Phys. Rev. Lett} \textbf{\bibinfo{volume}{31}},
  \bibinfo{pages}{1218} (\bibinfo{year}{1973}).

\bibitem[{\citenamefont{Ziesche and Lehmann}(1983)}]{ZL83}
\bibinfo{author}{\bibfnamefont{P.}~\bibnamefont{Ziesche}} \bibnamefont{and}
  \bibinfo{author}{\bibfnamefont{D.}~\bibnamefont{Lehmann}},
  \bibinfo{journal}{J. Phys. C} \textbf{\bibinfo{volume}{16}},
  \bibinfo{pages}{879} (\bibinfo{year}{1983}).

\bibitem[{\citenamefont{Lipavsk{\'y} et~al.}(2005)\citenamefont{Lipavsk{\'y},
  Morawetz, Kol{\'a}{\v c}ek, Mare{\v s}, Brandt, and Schreiber}}]{LMKMBSb04}
\bibinfo{author}{\bibfnamefont{P.}~\bibnamefont{Lipavsk{\'y}}},
  \bibinfo{author}{\bibfnamefont{K.}~\bibnamefont{Morawetz}},
  \bibinfo{author}{\bibfnamefont{J.}~\bibnamefont{Kol{\'a}{\v c}ek}},
  \bibinfo{author}{\bibfnamefont{J.~J.} \bibnamefont{Mare{\v s}}},
  \bibinfo{author}{\bibfnamefont{E.~H.} \bibnamefont{Brandt}},
  \bibnamefont{and}
  \bibinfo{author}{\bibfnamefont{M.}~\bibnamefont{Schreiber}},
  \bibinfo{journal}{Phys. Rev. B} \textbf{\bibinfo{volume}{71}},
  \bibinfo{pages}{024526} (\bibinfo{year}{2005}).

\bibitem[{\citenamefont{Lipavsk{\'y} et~al.}(2002)\citenamefont{Lipavsk{\'y},
  Kol{\'a}{\v c}ek, Morawetz, and Brandt}}]{LKMB02}
\bibinfo{author}{\bibfnamefont{P.}~\bibnamefont{Lipavsk{\'y}}},
  \bibinfo{author}{\bibfnamefont{J.}~\bibnamefont{Kol{\'a}{\v c}ek}},
  \bibinfo{author}{\bibfnamefont{K.}~\bibnamefont{Morawetz}}, \bibnamefont{and}
  \bibinfo{author}{\bibfnamefont{E.~H.} \bibnamefont{Brandt}},
  \bibinfo{journal}{Phys. Rev. B} \textbf{\bibinfo{volume}{66}},
  \bibinfo{pages}{134525} (\bibinfo{year}{2002}).

\bibitem[{\citenamefont{Kiejna and Wojciechowski}(1996)}]{KiWo96}
\bibinfo{author}{\bibfnamefont{A.}~\bibnamefont{Kiejna}} \bibnamefont{and}
  \bibinfo{author}{\bibfnamefont{K.~F.} \bibnamefont{Wojciechowski}},
  \emph{\bibinfo{title}{Metal surface electron physics}}
  (\bibinfo{publisher}{Elsevier Science}, \bibinfo{address}{Oxford},
  \bibinfo{year}{1996}).

\bibitem[{\citenamefont{Wadati and Iida}(2007)}]{WI07}
\bibinfo{author}{\bibfnamefont{M.}~\bibnamefont{Wadati}} \bibnamefont{and}
  \bibinfo{author}{\bibfnamefont{T.}~\bibnamefont{Iida}},
  \bibinfo{journal}{Phys. Lett. A} \textbf{\bibinfo{volume}{360}},
  \bibinfo{pages}{423} (\bibinfo{year}{2007}).

\bibitem[{\citenamefont{Squire and March}(2007)}]{SM07}
\bibinfo{author}{\bibfnamefont{R.~H.} \bibnamefont{Squire}} \bibnamefont{and}
  \bibinfo{author}{\bibfnamefont{N.~H.} \bibnamefont{March}},
  \bibinfo{journal}{Intern. J. Quant. Chem.} \textbf{\bibinfo{volume}{107}},
  \bibinfo{pages}{3013} (\bibinfo{year}{2007}).

\bibitem[{\citenamefont{Care and March}(1975)}]{CM75}
\bibinfo{author}{\bibfnamefont{C.~M.} \bibnamefont{Care}} \bibnamefont{and}
  \bibinfo{author}{\bibfnamefont{N.~H.} \bibnamefont{March}},
  \bibinfo{journal}{Adv. Phys.} \textbf{\bibinfo{volume}{24}},
  \bibinfo{pages}{101} (\bibinfo{year}{1975}).

\end{thebibliography}

\end{document}